\documentclass[aps,prl,showpacs,twocolumn,groupedaddress]{revtex4}
\usepackage{graphicx}

\begin{document}

\title{Lattice analysis for the energy scale of QCD phenomena}

\author{Arata~Yamamoto}
\affiliation{Department of Physics, Faculty of Science, Kyoto University, Kitashirakawa, Sakyo, Kyoto 606-8502, Japan}

\author{Hideo~Suganuma}
\affiliation{Department of Physics, Faculty of Science, Kyoto University, Kitashirakawa, Sakyo, Kyoto 606-8502, Japan}

\date{\today}

\begin{abstract}
We formulate a new framework in lattice QCD to study the relevant energy scale of QCD phenomena.
By considering the Fourier transformation of link variable, we can investigate the intrinsic energy scale of a physical quantity nonperturbatively.
This framework is broadly available for all lattice QCD calculations.
We apply this framework for the quark-antiquark potential and meson masses in quenched lattice QCD.
The gluonic energy scale relevant for the confinement is found to be less than 1 GeV in the Landau or Coulomb gauge.
\end{abstract}


\maketitle

Physics covers a huge number of phenomena on various scales, i.e., from elementary particles to the universe.
Nature is described in different ways depending on the energy scale.
The energy scale characterizes physical phenomena and is important in understanding them.

The classical QCD Lagrangian is scale invariant, except for quark mass terms.
However, quantum effects yield the scale anomaly and violate the scale invariance, and the energy scale appears in the real world.
The appearance of the energy scale enriches QCD phenomena and creates their hierarchy.
In high energy, due to the weak coupling or asymptotic freedom, the phenomena are governed by perturbative QCD \cite{Gr73,Po73}.
In contrast, in low energy, there are complicated and important nonperturbative phenomena, i.e., quark confinement, chiral symmetry breaking, and so on \cite{Gr03,Ha99,Ma87}.
Many of the QCD phenomena would have their intrinsic energy scale.
These phenomena originate from a single Lagrangian but seem like different physics.
The knowledge about the energy scale of QCD phenomena would be useful in understanding complicated QCD dynamics.

One possible way to study the intrinsic energy scale is to calculate a physical quantity after cutting a part of the region of momentum space artificially.
From the relation between the cut region and the resulting quantity, we can understand the relevant energy scale for the quantity.
This concept is used, for example, in the Swinger-Dyson approach \cite{Ii05}.
In this letter, we introduce this concept to lattice QCD, which is the nonperturbative and first-principle calculation \cite{Cr81,Ro92}.
The fundamental degree of freedom in lattice QCD is the link variable, which represents the gluon on lattice.
We cut a part of the link variable in momentum space, and investigate its effect on the resulting quantity.


The procedure is formulated as the following five steps.

1.~We generate a gauge configuration, i.e., link variables $U_{\mu}(x)(\in {\rm SU(3)})$, by the usual Monte Carlo simulation in lattice QCD.
Since the procedure is not gauge independent, we fix the configurations with some gauge.
It is desirable to choose a physically meaningful gauge on lattice, such as the Landau gauge, which has the corresponding continuum theory and the minimum fluctuation of the gauge field.
In this letter, we mainly use the Landau gauge for numerical calculation, unless otherwise stated.

2.~Using discrete Fourier transformation in the 4-dimensional Euclid space, we construct the momentum-space link variable ${\tilde U}_{\mu}(p)$ as
\begin{eqnarray}
{\tilde U}_{\mu}(p)=\frac{1}{L^4}\sum_x U_{\mu}(x)\exp(i {\textstyle \sum_\nu} p_\nu x_\nu),
\end{eqnarray}
where $L$ is a number of lattice sites in one direction.
When numerical simulation is performed in a certain lattice size with periodic boundary condition, the corresponding momentum space is the lattice with the same structure.
The first Brillouin zone of the momentum space is a hypercube with each side of ($-\pi/a, \pi/a$].

\begin{figure}[t]
\begin{center}
\includegraphics[scale=0.5]{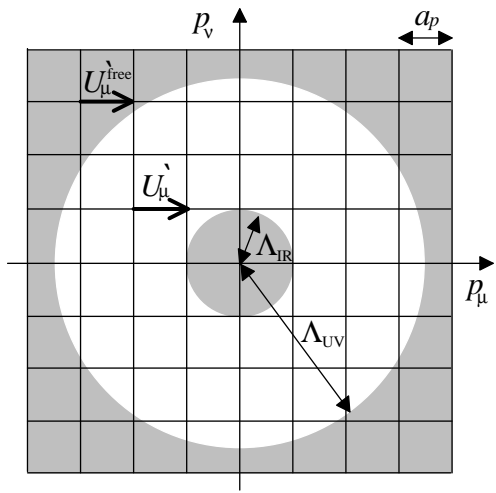}
\caption{\label{fig1}
The two-dimensional schematic figure of the UV cut $\Lambda_{\rm UV}$ and the IR cut $\Lambda_{\rm IR}$ on momentum-space lattice.
The momentum-space link variable ${\tilde U}_{\mu}(p)$ is replaced with the free one ${\tilde U}^{\rm free}_{\mu}(p)$ in the cut (shaded) regions.
$a_p$ is the momentum-space lattice spacing.
}
\end{center}
\end{figure}

3.~We impose a ``cut" on ${\tilde U}_{\mu}(p)$ in some region of the momentum space, for example, a ultraviolet (UV) cut or a infrared (IR) cut.
Outside the cut, the link variable is replaced with the free link variable ${\tilde U}^{\rm free}_{\mu}(p)=\delta_{p0}$.
In this letter, we estimate the energy scale by the 4-momentum length $\sqrt{p^2}=\sqrt{\sum_\mu p_\mu p_\mu}$, and define the momentum-space link variable ${\tilde U}_{\mu}^{\Lambda}(p)$ with the UV cut $\Lambda_{\rm UV}$ and the IR cut $\Lambda_{\rm IR}$ as
\begin{equation}
\label{eq2}
{\tilde U}_{\mu}^{\Lambda}(p)= \Bigg\{
\begin{array}{cc}
{\tilde U}_{\mu}(p) & (\Lambda_{\rm UV} \ge \sqrt{p^2} \ge \Lambda_{\rm IR})\\
\delta_{p0} & (\sqrt{p^2} > \Lambda_{\rm UV} \ {\rm or} \ \sqrt{p^2} < \Lambda_{\rm IR})
\end{array}
\end{equation}
These cuts are schematically depicted in Fig.~\ref{fig1}.
Of course, we can consider other kinds of cuts, for example, the cut for the intermediate-momentum region, the anisotropic cut, and so on.

4.~We reconstruct the coordinate-space link variable $U^{\Lambda}_{\mu}(x)$ with the cut by the inverse Fourier transformation.
Since the naive inverse Fourier transformation of ${\tilde U}_{\mu}^{\Lambda}(p)$ is not a SU(3) matrix, we define $U^{\Lambda}_{\mu}(x)(\in {\rm SU(3)})$ by maximizing
\begin{eqnarray}
{\rm ReTr}[U^{\Lambda}_{\mu}(x)^{\dagger}U'_{\mu}(x)],
\end{eqnarray}
where
\begin{eqnarray}
U'_{\mu}(x)=\sum_p {\tilde U}_{\mu}^{\Lambda}(p)\exp(-i {\textstyle \sum_\nu} p_\nu x_\nu).
\end{eqnarray}

5.~Using this link variable $U^{\Lambda}_{\mu}(x)$, we compute physical quantities in just the same way as the original SU(3) lattice QCD.

By this procedure, we can investigate what energy scale of the gluon is essential for a QCD quantity, nonperturbatively and quantitatively.
We have only to replace $U^{\Lambda}_{\mu}(x)$ instead of $U_{\mu}(x)$ in the usual lattice QCD calculation.
Therefore this framework can be applied to whatever can be calculated by lattice QCD, and would be widely useful for QCD phenomena.

We would like to apply this procedure to several fundamental quantities in quenched lattice QCD.
The numerical simulation is performed with the $\beta=6.0$ plaquette gauge action on $16^4$ lattice.
The lattice spacing $a$ is about 0.10 fm \cite{Ya08}.
Then the corresponding momentum-space lattice is also $16^4$ lattice, and the momentum-space lattice spacing $a_p=2\pi/La$ is about 0.77 GeV.

First, we calculate the quark-antiquark ($Q\bar Q$) potential with the UV or IR cut.
The $Q\bar Q$ potential is extracted from the expectation value of the Wilson loop.
To enhance the ground-state component, the smearing method is used for the spatial link variables in the Wilson loop \cite{Al87}.
The configuration number is 50, but effectively large by averaging all the parallel-translated Wilson loops in one configuration.

The physical $Q\bar Q$ potential is written with a Coulomb plus linear potential as
\begin{eqnarray}
V(R)=\sigma R -\frac{A}{R} +C,
\label{VQQ}
\end{eqnarray}
where $R$ is the distance between the quark and the antiquark \cite{Cr7980}.
The Coulomb potential is explained by the perturbative one-gluon exchange, and its coefficient $A$ is about 0.26.
The linear confinement potential is purely a nonperturbative phenomenon.
Its coefficient, or the string tension, $\sigma$ is about 0.89 GeV/fm, and $\sigma a^2\simeq 0.051$ in $\beta=6.0$ lattice unit.
The constant term $C$ is physically irrelevant and depends on the regularization.

\begin{figure}[t]
\begin{center}
\includegraphics[scale=1.2]{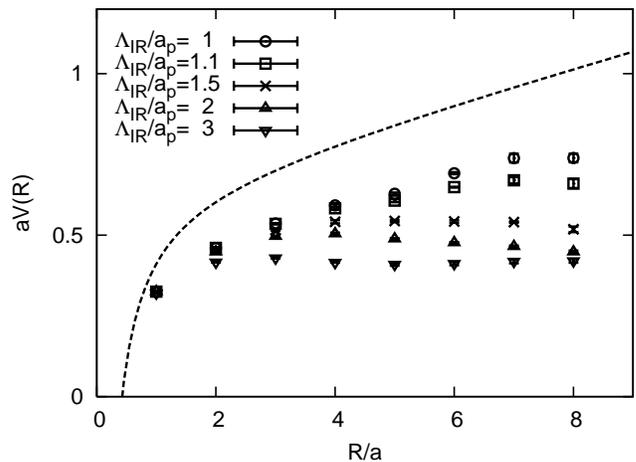}
\caption{\label{fig2}
The $Q\bar Q$ potential $V(R)$ with the IR cut $\Lambda_{\rm IR}$ against the interquark distance $R$.
The unit is scaled with the lattice spacings $a\simeq 0.10$ fm and $a_p \simeq 0.77$ GeV.
The broken line is the original $Q\bar Q$ potential in lattice QCD.
}
\end{center}
\end{figure}

We show the $Q\bar Q$ potential $V(R)$ with the IR cut $\Lambda_{\rm IR}$ in Fig.~\ref{fig2}.
When $\Lambda_{\rm IR}/a_p=1$ and 1.1, the string tension slightly decreases from the original $Q\bar Q$ potential.
When $\Lambda_{\rm IR}/a_p\ge 1.5$, the linear confinement potential completely disappears and the $Q\bar Q$ potential becomes only the perturbative Coulomb potential.
In Table \ref{tab1}, we list the asymptotic string tension $\sigma_{\rm asym}$ which is estimated by fitting the $Q\bar Q$ potential in $4\le R/a \le 8$ with a linear function $\sigma_{\rm asym}R + {\rm const}$.
The asymptotic string tension is approximately zero in $\Lambda_{\rm IR}/a_p\ge 1.5$.
This result means that the $Q\bar Q$ confinement potential is made from the low-momentum gluon, and that its physical energy scale is less than 1 GeV in the Landau gauge. 

\begin{table}[b]
\caption{\label{tab1}
The asymptotic string tension $\sigma_{\rm asym}$ with the IR cut $\Lambda_{\rm IR}$.
The results of the Landau gauge and the Coulomb gauge are shown.
In original lattice QCD with $\beta =6.0$, the string tension $\sigma a^2$ is about 0.051.
}
\begin{ruledtabular}
\begin{tabular}{ccc}
$\Lambda_{\rm IR}/a_p$ &$\sigma_{\rm asym}a^2$ (Landau)& $\sigma_{\rm asym}a^2$ (Coulomb)\\
\hline
1.0 & 0.0469(58) & 0.0289(58) \\
1.1 & 0.0311(49) & 0.0190(59)\\
1.5 & -0.0019(20) & 0.0024(25)\\
2.0 & -0.0132(6) & -0.0041(10) \\
3.0 & 0.0003(12) & 0.0058(10) \\
\end{tabular}
\end{ruledtabular}
\end{table}

Since the procedure is not gauge invariant, it is desired to check the gauge dependence of the resulting energy scale among several reasonable gauge choices in the step 1.
In Table \ref{tab1}, we also show the result of the Coulomb gauge for the gauge choice.
The value of $\sigma_{\rm asym}$ is different between the Coulomb gauge and the Landau gauge.
However, the qualitative behavior is the same, and the energy scale for the confinement potential is less than about 1 GeV also in the Coulomb gauge.

\begin{figure}[b]
\begin{center}
\includegraphics[scale=1.2]{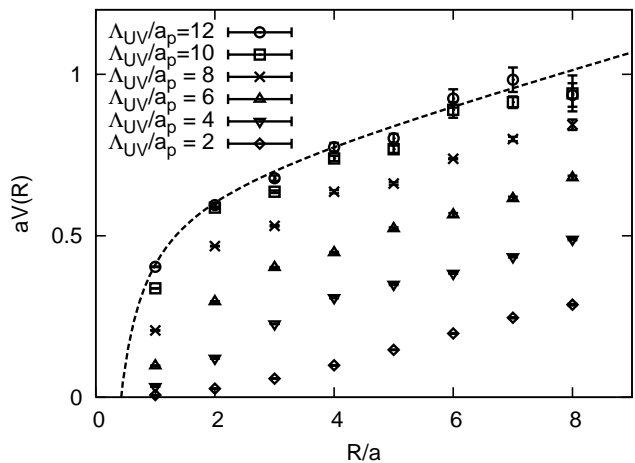}
\caption{\label{fig3}
The $Q\bar Q$ potential with the UV cut $\Lambda_{\rm UV}$.
The notation is the same as Fig.~\ref{fig2}.
}
\end{center}
\end{figure}

The $Q\bar Q$ potential with the UV cut $\Lambda_{\rm UV}$ is shown in Fig.~\ref{fig3}.
As the UV cut becomes lower, the Coulomb potential gradually weakens.
In contrast, the string tension remains unchanged regardless of the UV cut.
As for the constant term $C$ in Eq.~(\ref{VQQ}), it is considered to be mainly given by the lattice regularization of the Coulomb singularity, that is, a perturbative origin.
Actually, the constant term becomes smaller with lower UV cut.
With the very low UV cut, the $Q\bar Q$ potential is almost the linear potential, and the Coulomb coefficient and the constant term is nearly zero or slightly negative.
For example, with $\Lambda_{\rm UV}/a_p=2$, the fitting result by Eq.~(\ref{VQQ}) is $\sigma a^2=0.0459(15)$, $A=-0.056(5)$, and $Ca=-0.095(7)$.

\begin{figure}[t]
\begin{center}
\includegraphics[scale=1.2]{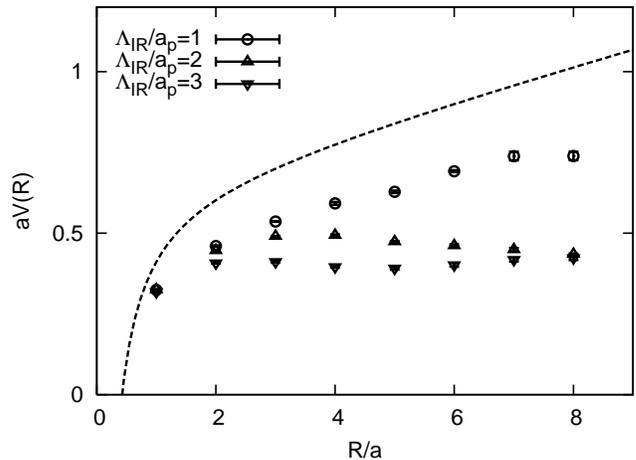}
\caption{\label{fig4}
The $Q\bar Q$ potential $V(R)$ with the IR cut $\Lambda_{\rm IR}$ by ${\rm max}(p_1,p_2,p_3,p_4)$ instead of $\sqrt{p^2}$ in Eq.~(\ref{eq2}).
The notation is the same as Fig.~\ref{fig2}.
}
\end{center}
\end{figure}

Now we consider another type of the cut, i.e., the cut by ${\rm max}(p_1,p_2,p_3,p_4)$ instead of $\sqrt{p^2}$ in Eq.~(\ref{eq2}).
This type of the cut forms a 4-dimensional hypercube, and respects the momentum-space lattice structure.
We find that the $Q\bar Q$ potential with this type of the cut displays the same qualitative behavior as before in both the UV and IR case, and gives almost the same energy scale.
A part of the results is shown in Fig.~\ref{fig4}.

Next, we apply this framework to the meson mass calculation.
The quark propagator is calculated with the clover fermion action, which is the $O(a)$-improved Wilson fermion action.
The hopping parameters are $\kappa=0.1200$, 0.1300, and 0.1340, and the corresponding pion masses are about 2.9 GeV, 1.8 GeV, and 1.3 GeV, respectively \cite{Ya08}.
The configuration number is 100 here.

\begin{figure}[t]
\begin{center}
\includegraphics[scale=1.2]{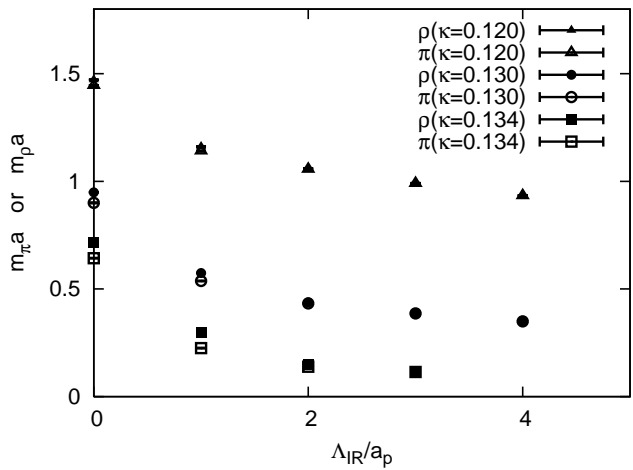}
\caption{\label{fig5}
Pion mass $m_\pi$ and $\rho$-meson mass $m_\rho$ with the IR cut $\Lambda_{\rm IR}$.
The results of three hopping parameters $\kappa=0.1200$, 0.1300, and 0.1340 are shown.
The unit is scaled with the lattice spacings $a\simeq 0.10$ fm and $a_p \simeq 0.77$ GeV.
The values at $\Lambda_{\rm IR}/a_p=0$ is calculated in original lattice QCD \cite{Ya08}.
}
\end{center}
\end{figure}

Figure \ref{fig5} shows pion mass $m_\pi$ and $\rho$-meson mass $m_\rho$ with the IR cut $\Lambda_{\rm IR}$.
When more IR region is cut, the pion mass and the $\rho$-meson mass become smaller.
In addition, the pion and $\rho$-meson masses degenerate in $\Lambda_{\rm IR}/a_p\ge 2$.
This energy scale is consistent with the energy scale of the disappearance of the $Q\bar Q$ confinement potential.
Therefore this degeneracy is considered to suggest that the quark and the antiquark become unbound and quasi-free and that $m_\pi$ and $m_\rho$ are equal to the sum of the quark and antiquark masses.
We have also calculated these meson masses with the UV cut, and found that these masses decrease also by the UV cut but the degeneracy does not occur.
It would be also interesting to investigate the meson masses and the chiral condensate by this procedure with the staggered fermion or the overlap fermion.

In summary, we have proposed the new framework in lattice QCD to investigate the relation between a QCD phenomenon and its typical energy scale.
Using Fourier transformation, we construct momentum-space link variables, cut a part of them in momentum space, and reconstruct coordinate-space link variables with the cut.
By lattice QCD calculations with this link variables, we can investigate the gluonic energy scale of QCD quantities.
Since what is needed is only the replacement of the link variables, this is easily available for all lattice QCD calculations.
In addition, this kind of framework would be broadly applied to not only lattice QCD but also any discretized theory.
By such a framework, we can systematically analyze the momentum structure of physical phenomena.

We have applied this framework to the calculation for the $Q\bar Q$ potential and meson masses.
When IR gluons are cut, the nonperturbative confinement potential vanishes in the $Q\bar Q$ potential, and pion and $\rho$-meson degenerate and seem to be the quasi-free quark and antiquark pair.
The relevant gluonic energy scale for quark confinement is less than about 1 GeV in the Landau gauge and the Coulomb gauge.

A.~Y.~and H.~S.~are supported by a Grant-in-Aid for Scientific Research [(C) No.~20$\cdot$363 and (C) No.~19540287] in Japan.
This work is supported by the Global COE Program, ``The Next Generation of Physics, Spun from Universality and Emergence".
The lattice QCD calculations are done on NEC SX-8R at Osaka University.

\end{document}